\documentclass[page-classic]{epl2}

\makeatletter
\usepackage{pxfonts}

\begin{document}

\title{Lorentz Transform of Black Body Radiation Temperature}

\shorttitle{Black Body Radiation Temperature} 

\institute{                    
   Fukui Prefectural University, 910-1195 Fukui, JAPAN
}
 

\author{Tadas K Nakamura}

\abstract{
The Lorentz transform of black body radiation has been investigated
from the view point of relativistic statistical mechanics. The result
shows that the well known expression with the directional temperature
can be derived based on the inverse temperature four vector. The directional
temperature in the past literature was the result of mathematical
manipulation and its physical meaning is not clear. The inverse temperature
four vector has, in contrast, clear meaning to understand relativistic
thermodynamical processes.
}

\maketitle

\section{Introduction}

It is well known that black body radiation {obeys} 
the Planck distribution;
the following expression can be found in textbooks:\begin{equation}
n(\omega)\, d\omega=\frac{\omega^{2}}{2\pi^{2}[\exp(\omega/T)-1]}\, d\omega\,,\label{eq:planck}\end{equation}
where $n$ is the number density of photons with frequency $\omega$
and $T$ is the temperature. (The unit system is such that light speed
= Boltzmann constant = Planck constant =1.) The above formula is the
expression in the reference frame of the black body cavity. The expression
for an observer moving relative to the cavity has been calculated
in the context of the cosmic microwave background (CMB). 

If the solar system is moving relative to the CMB's rest frame, the
distribution of CMB observed at the earth will be different from (\ref{eq:planck})
and the difference can tell us the motion of the solar system. Several
authors published the same result in the same year \cite{bracewell-conklin68,henry-etal68,peebles-wilkinson68};
the number density of photons coming from the solid angle $\Omega$
with frequency $\omega$ is expressed as \begin{equation}
n(\omega,\Omega)\, d\omega d\Omega=\frac{\omega^{2}}{2\pi^{2}[\exp(\omega/T_{\mathrm{eff}}(\theta))-1]}\, d\omega d\Omega\,.\label{eq:henry}\end{equation}
In the above expression $T_{\mathrm{eff}}$ is the effective temperature,
which is called {}``directional temperature'', defined as\begin{equation}
T_{\mathrm{eff}}(\theta)=\frac{T_{*}\sqrt{1-V^{2}}}{1-V\cos(\theta)}\,,\label{eq:Teff}\end{equation}
where $T_{*}$ is the black body temperature as measured in the cavity
rest frame {[}$T$ in (\ref{eq:planck})], $V$ is the observer's
velocity, and $\theta$ is the angle between the observer's motion
and the direction of observation.

The above result has been obtained from a purely mathematical manipulation,
and no thermodynamical consideration is used to derive it. Therefore,
$T_{\mathrm{eff}}$ is just a mathematical shorthand in the formula
and it is not clear whether $T_{\mathrm{eff}}$ has the meaning of
temperature or not. This is sufficient for its original purpose, that
is, to determine the motion of the solar system relative to CMB (e.g.,
\cite{kogut-etal93ApJ}). However, it is not enough when we wish to
investigate the thermodynamical properties of the black body radiation.
For example, suppose 
{a moving matter that is immersed into a black body radiation.} When
the matter has {proper temperature (temperature measured in
the comoving
frame) $T$} such that $T_{\mathrm{eff}}(0)<T<T_{\mathrm{eff}}(\pi/2)$,
then we cannot tell the direction of heat flow; usually heat flows
from the higher temperature to the lower temperature, but {from knowing
$T_{\mathrm{eff}}$ it cannot be determined} which has the higher temperature.

An attempt to understand the thermodynamics of moving black body radiation
has been made in relatively recent years. Aldrovandi and Gariel \cite{aldrovandi-gariel92}
regarded (\ref{eq:henry}) as the temperature transformation law and
{concluded that} the temperature of a moving object becomes higher. Costa
and Matsas \cite{costa-matsas95} calculated the photon distribution
using the Unruh-DeWitt detector moving relative to the radiation,
and showed the equilibrium distribution does not have the form of
(\ref{eq:planck}). Form this fact, Landsberg and Matsas \cite{landsberg-costa96PhLA,landsberg-matsas04PhyA}
claimed that the relativistic temperature transformation is impossible
and the concept of temperature can be defined only in the comoving
reference frame. Ares de Parga et al., \cite{aresdeparga05JPhA} have
examined this problem based on the theory they have proposed, and
{concluded that}
the expression with the directional temperature can be understood
within their theory.

What we would like to show in the present letter is that the formula
(\ref{eq:henry}) can be consistently derived from the view point
of relativistic statistical mechanics. The directional temperature
in (\ref{eq:Teff}) is just a shorthand notation, and does not have
thermodynamical implication as a temperature. In contrast, the calculation
here derives the same expression as (\ref{eq:Teff}) based on the
inverse temperature four vector proposed in the context of relativistic
thermodynamics. This inverse temperature four vector has clear thermodynamical
meaning because it comes from the conservation law of energy-momentum,
just in the same way as the inverse temperature in conventional non-relativistic
statistical mechanics.

The inverse temperature four vector was originally introduced by van-Kampen
\cite{kampen68PhRv} in the controversy on relativistic thermodynamics
in 1960s, and later refined by Israel in a more transparent form \cite{Israel76AnPhy}.
There are a number of different formulations of relativistic thermodynamics
(see, e.g., \cite{yuen70AmJPh}), however, it can be shown that
the other formulations can be derived from the van Kampen-Israel theory
with the inverse temperature four vector \cite{tadas08arXiv}. We
will see in the present paper that the inverse temperature four vector
can be also applied for the covariant treatment of black body radiation.

\section{Momentum in Statistical Mechanics}

Before exploring the black body radiation in relativity, we briefly
demonstrate our tactics with a simple example of non-relativistic
classical ideal gas. Let $f(\mathbf{p})$ be the single particle distribution
function as a function of momentum $\mathbf{p}$. We employ the maximum
entropy approach (e.g., \cite{jaynes83}) to calculate the equilibrium
distribution, which is obtained by maximizing the following entropy\begin{equation}
S=\int f\ln f\, d\mathbf{p}\,,\label{eq:entropy}\end{equation}
under the following constraints of the particle number and energy
conservation,\begin{eqnarray}
\int f(\mathbf{p})\, d\mathbf{p}=\textrm{1,}\;\; N\int\frac{1}{2m}\mathbf{p}^{2}f(\mathbf{p})\, d\mathbf{p} & = & \textrm{total energy,}\label{eq:energy}\end{eqnarray}
where $m$ is the mass of a particle and $N$ is the total number
of particles. The equilibrium distribution is obtained as \begin{equation}
f(\mathbf{p})\propto\exp\left(\alpha-\frac{\beta\mathbf{p}^{2}}{2m}\right)\,.\label{eq:mawell}\end{equation}
The parameter $\alpha$ and $\beta$ in the above expression are the
Lagrange's coefficients arising from the constraints of (\ref{eq:energy}),
and should be determined appropriately to satisfy the constraints. 

When we observe this distribution from a frame moving with the relative
velocity $\mathbf{V}$, then the Galilei transform of the distribution
function can be calculated by replacing $\mathbf{p}\rightarrow\mathbf{p}-m\mathbf{V}$
as\begin{equation}
f(\mathbf{p})\propto\exp\left(\alpha-\frac{\beta}{2m}(\mathbf{p}-m\mathbf{V})^{2}\right)\,.\label{eq:maxwell2}\end{equation}
This expression is obtained by mathematical transform of (\ref{eq:mawell})
and no thermodynamical consideration is required once (\ref{eq:mawell})
is given; this corresponds to the derivation of (\ref{eq:henry})
\cite{bracewell-conklin68,henry-etal68,peebles-wilkinson68}. 

There can be another way to derive (\ref{eq:maxwell2}) from the view
point of entropy maximization. We introduce another conservation law,
the momentum conservation namely, in addition to the energy constraint
of (\ref{eq:energy}):\begin{equation}
N\int\mathbf{p}f(\mathbf{p})\, d\mathbf{p}=\textrm{total momentum\,.}\label{eq:momentum}\end{equation}
Then three other Lagrange's coefficients appear corresponding to the
three components of momentum, and the distribution becomes\begin{equation}
f(\mathbf{p})\propto\exp\left(\alpha'-\frac{\beta'\mathbf{p}^{2}}{2m}+\beta'_{x}p_{x}+\beta'_{y}p_{y}+\beta'_{z}p_{z}\right)\,.\end{equation}
The Lagrange's coefficients $\alpha$', $\beta$', $\beta'_{i}$($i=x,y,z$)
should be determined to satisfy the constraints (\ref{eq:energy})
and (\ref{eq:momentum}). Since this distribution has the same energy
and momentum as (\ref{eq:maxwell2}), the coefficients are\begin{eqnarray}
\alpha' & = & \alpha+\frac{1}{2}m\mathbf{V}^{2}\,,\nonumber \\
\beta' & = & \beta\,,\label{eq:invtemps}\\
\beta'_{i} & = & \beta V_{i}\;(i=x,y,z)\,,.\nonumber \end{eqnarray}

The above result has something more than the derivation of (\ref{eq:maxwell2}).
Since the coefficients $\beta_{i}$ are obtained in a same way as
to derive the inverse temperature $\beta$, they have similar meaning
in thermodynamics. If two bodies with different temperature are thermally
connected, in other words, there is random energy exchange between
the bodies, the energy flow is such as to reduce the difference of
the inverse temperature $\beta$. We can generalize this statement
to the random momentum exchange between two bodies moving relative
to each other. The momentum is transferred in the direction to reduce
the relative velocity because it increases the total entropy. The result
is the frictional force between the two bodies.

The example in this section demonstrates the role of momentum as a
thermodynamical parameter; three more inverse temperature arise corresponding
to the three components of momentum. This non-relativistic example
may be rather trivial because the inverse temperature of energy ($\beta$)
is unchanged under the Galilei transform. However, if we wish to construct
covariant relativistic thermodynamics, not only energy but also the
three components of momentum should be regarded as thermodynamical
quantities because they are components of a four vector. Correspondingly
inverse temperatures $\beta$ and $\beta_{i}$ ($i=x,y,z$) forms
a four vector when we generalize the calculation to relativity. This
four vector is the covariant expression of the inverse temperature
in the relativistic thermodynamics proposed by van Kampen and Israel \cite{kampen68PhRv,Israel76AnPhy}.

It is possible to perform the same calculation as above for a relativistic
ideal gas, however, there is a subtle point in the maximum entropy
calculation, and controversy is still going on \cite{dunkel07,tadas09rel-eq}.
Therefore, we dare not examine this subject in the present letter
and move onto the black body radiation in the following.

\section{Black Body Temperature}

What we have learned form the previous section is that the Boltzmann
factor $\exp(-\beta E)$ should be replaced with \begin{equation}
\exp(-\beta E)\rightarrow\exp\left(-\sum_{\mu=0}^{3}\beta_{\mu}P^{\mu}\right)\,,\end{equation}
where $P^{\mu}$ is the energy-momentum of the system, to include
the momentum as a thermodynamical quantity. As we have seen in the
previous section, the inverse temperatures $\beta_{\mu}$ are obtained
from statistical mechanics, and thus we know their roles in thermodynamics.
However, we do not know whether they are transformed as a four vector
or not at this stage; we used the notation with $\Sigma$ to emphasize
this point in the above expression.

In the case of a photon gas, the energy-momentum of a photon is given
by its four dimensional wave number. Suppose a photon gas in a black
body cavity, which is moving in the $z\,(=x^{3})$ direction with
the velocity $V$ in one reference frame. The $x$ and $y$ components
of the momentum vanish 
{because of the symmetry,} so we can set $\beta_{x}=\beta_{y}=0$.
Following the standard procedure in the statistical mechanics, the
number of photons in one wave mode is obtained as \begin{equation}
N_{i}=\frac{{\displaystyle 1}}{{\displaystyle \exp(\beta_{t}\omega_{i}-\beta_{z}k_{iz})}-1}\,,\end{equation}
where $\omega_{i}$ and $k_{iz}$ are the frequency and wave number
of the $i$-th mode. 

Let us introduce polar coordinates $(r,\theta,\phi)$to calculate waves
propagating in one direction; we choose
the coordinates such that $\theta=0$ is the direction of the spatial
wave vector. The number of photons in the limit of continuous frequency
can be calculated in the same way as to derive (\ref{eq:planck}),
which yields\begin{equation}
n(\omega,\Omega)\, d\omega d\Omega=\frac{\omega^{2}}{2\pi^{2}[\exp[(\beta_{t}-\beta_{z}\cos\theta)\omega]-1]}\, d\omega d\Omega\,.\label{eq:henry2}\end{equation}
We have used the dispersion relation of photons $\omega^{2}=k_{z}^{2}$
to obtain the above expression. The inverse temperature $\beta_{\mu}$
should be determined such that the distribution $n(\omega,\Omega)$
gives the total energy-momentum correctly. This can be done in the
same way as we have done in deriving (\ref{eq:invtemps}). Comparing
(\ref{eq:henry}) and (\ref{eq:henry2}) we obtain \begin{equation}
\beta_{t}=\frac{1}{T_{*}\sqrt{1-V^{2}}}\,,\;\;\beta_{z}=\frac{V}{T_{*}\sqrt{1-V^{2}}}\,.\end{equation}

This result can be generalized to a covariant form as\begin{equation}
\beta_{\mu}=\frac{u_{\mu}}{T_{*}}\,,\end{equation}
where $u_{\mu}$ is the relative four velocity between the radiation
and the observer. We understand the inverse temperatures $\beta_{\mu}$
form a four vector from the above explicit form. This is the inverse
temperature four vector in the van Kampen-Israel theory \cite{kampen68PhRv,Israel76AnPhy}.
Unlike the directional temperature $T_{\mathrm{eff}}$, the above
four vector $\beta_{\mu}$ has been derived from the view point of
statistical mechanics. Therefore $\beta_{\mu}$ has clear meaning
as inverse temperatures, and can tell the direction of 
{the thermal energy-momentum
exchange as being discussed} by van Kampen \cite{kampen68PhRv}.

\section{Concluding Remarks}

Brief remarks are to be made on the past literature before closing
this letter. Aldrovandi \cite{aldrovandi-gariel92} examined the temperature
transformation assuming the directional temperature has the thermodynamical
meaning somehow; he did not give the reason for this assumption as
he states {}``we prefer to avoid an 'inside' thermodynamical discussion.''
Ares de Parga et al., \cite{aresdeparga05JPhA} have their own reasoning
to interpret the directional temperature within the theory they proposed.
The two results seems to contradict each other: the former suggests
the higher temperature for a moving body whereas the latter predicts
it lower. This contradiction is quite similar to the controversy on
the relativistic thermodynamics in 1960s. Both may be consistent within
each framework, however, the author of this paper believes the expression
with inverse temperature four vector is {clear and transparent.}

\bigskip

Costa and Matsas \cite{costa-matsas95} have calculated the particle
distribution of photons using the Unruh-DeWitt detector and obtained
the following expression,\begin{equation}
n(\omega)=\frac{T_{*}\sqrt{1-v^{2}}}{4\pi v}\,\ln\left(\frac{1-\exp(-\omega\sqrt{1+v}/T_{0}\sqrt{1-v})}{1-\exp(-\omega\sqrt{1-v}/T_{0}\sqrt{1+v})}\right)\,.\label{eq:costa}\end{equation}
The expression can be also obtained by integrating (\ref{eq:henry})
over the solid angle; this can be understood because the Unruh-DeWitt
detector measures energy only and directional dependence is smeared
out. The above expression does not have the form of the Planck distribution,
and from this fact Landsberg and Matsas \cite{landsberg-costa96PhLA,landsberg-matsas04PhyA}
argued that{Lorentz transforming the temperature would be impossible.}

However, a distribution function changes its shape in general when
expressed as a function of energy only. For example, suppose a non-relativistic
gas with distribution function (\ref{eq:maxwell2}). When we express
the distribution as a function of energy ignoring directional dependence,
we obtain \begin{equation}
f(E)=\int f(\mathbf{p})\,\delta\left(E-\frac{1}{2m}\mathbf{p}^{2}\right)\, d\mathbf{p}=\frac{T}{V_{0}\sqrt{2\pi E}}\sinh\left(\frac{mV_{0}^{2}}{T}\right)\exp\left(-\frac{E}{T}\right)\,,\end{equation}
which does not have the form of Boltzmann distribution $f\propto\exp(-\beta E)/\sqrt{E}$;
obviously this does not mean the concept of temperature $T$ is invalid
in the Galilei transform.

Similarly the concept of temperature (or inverse temperature) is still
valid even when (\ref{eq:costa}) does not have the form of the Planck
distribution. What Landsberg and Matsas \cite{landsberg-costa96PhLA,landsberg-matsas04PhyA}
argued should be understood that the Lorentz transform of temperature
is impossible when one tries to express the temperature by a single
value; it is possible when we treat the inverse temperature as a four
vector. The present letter has shown that we can interpret the distribution
($\ref{eq:henry}$) as the statistical equilibrium state with the
inverse temperature four vector $\beta_{\mu}$ in the van Kampen-Israel
theory \cite{kampen68PhRv,Israel76AnPhy}.

\bibliographystyle{elsarticle-num}
\bibliography{rel-thermo}

\begin{thebibliography}{10}
\expandafter\ifx\csname url\endcsname\relax
  \def\url#1{\texttt{#1}}\fi
\expandafter\ifx\csname urlprefix\endcsname\relax\def\urlprefix{URL }\fi
\expandafter\ifx\csname href\endcsname\relax
  \def\href#1#2{#2} \def\path#1{#1}\fi

\bibitem{bracewell-conklin68}
T.~N. Bracewell, E.~K. Conklin, An observer moving in the 3 k radiation field,
  Nature 219 (1968) 1343--1344.
\newblock \href {http://dx.doi.org/10.1038/2191343a0}
  {\path{doi:10.1038/2191343a0}}.

\bibitem{henry-etal68}
G.~R. Henry, R.~B. Feduniak, J.~E. Silver, M.~A. Peterson, Distribution of
  blackbody cavity radiation in a moving frame of reference, Phys. Rev. 176~(5)
  (1968) 1451--1455.
\newblock \href {http://dx.doi.org/10.1103/PhysRev.176.1451}
  {\path{doi:10.1103/PhysRev.176.1451}}.

\bibitem{peebles-wilkinson68}
P.~J.~E. Peebles, D.~T. Wilkinson, Comment on the anisotropy of the primeval
  fireball, Phys. Rev. 174~(5) (1968) 2168.
\newblock \href {http://dx.doi.org/10.1103/PhysRev.174.2168}
  {\path{doi:10.1103/PhysRev.174.2168}}.

\bibitem{kogut-etal93ApJ}
A.~{Kogut}, C.~{Lineweaver}, G.~F. {Smoot}, C.~L. {Bennett}, A.~{Banday}, N.~W.
  {Boggess}, E.~S. {Cheng}, G.~{de Amici}, D.~J. {Fixsen}, G.~{Hinshaw}, P.~D.
  {Jackson}, M.~{Janssen}, P.~{Keegstra}, K.~{Loewenstein}, P.~{Lubin}, J.~C.
  {Mather}, L.~{Tenorio}, R.~{Weiss}, D.~T. {Wilkinson}, E.~L. {Wright},
  {Dipole Anisotropy in the COBE Differential Microwave Radiometers First-Year
  Sky Maps}, Astrophys J 419 (1993) 1--+.
\newblock \href {http://arxiv.org/abs/astro-ph/9312056}
  {\path{arXiv:astro-ph/9312056}}, \href {http://dx.doi.org/10.1086/173453}
  {\path{doi:10.1086/173453}}.

\bibitem{aldrovandi-gariel92}
R.~{Aldrovandi}, J.~{Gariel}, {On the riddle of the moving thermometers},
  Physics Letters A 170 (1992) 5--10.
\newblock \href {http://dx.doi.org/10.1016/0375-9601(92)90382-V}
  {\path{doi:10.1016/0375-9601(92)90382-V}}.

\bibitem{costa-matsas95}
S.~S. {Costa}, G.~E.~A. {Matsas}, {Temperature and relativity}, Physics Letters
  A 209 (1995) 155--159.
\newblock \href {http://arxiv.org/abs/gr-qc/9505045}
  {\path{arXiv:gr-qc/9505045}}, \href
  {http://dx.doi.org/10.1016/0375-9601(95)00843-7}
  {\path{doi:10.1016/0375-9601(95)00843-7}}.

\bibitem{landsberg-costa96PhLA}
P.~T. {Landsberg}, G.~E.~A. {Matsas}, {Laying the ghost of the relativistic
  temperature transformation}, Physics Letters A 223 (1996) 401--403.
\newblock \href {http://arxiv.org/abs/physics/9610016}
  {\path{arXiv:physics/9610016}}, \href
  {http://dx.doi.org/10.1016/S0375-9601(96)00791-8}
  {\path{doi:10.1016/S0375-9601(96)00791-8}}.

\bibitem{landsberg-matsas04PhyA}
P.~T. {Landsberg}, G.~E.~A. {Matsas}, {The impossibility of a universal
  relativistic temperature transformation}, Physica A Statistical Mechanics and
  its Applications 340 (2004) 92--94.
\newblock \href {http://dx.doi.org/10.1016/j.physa.2004.03.081}
  {\path{doi:10.1016/j.physa.2004.03.081}}.

\bibitem{aresdeparga05JPhA}
G.~{Ares de Parga}, B.~{L{\'o}pez-Carrera}, F.~{Angulo-Brown}, {A proposal for
  relativistic transformations in thermodynamics}, Journal of Physics A
  Mathematical General 38 (2005) 2821--2834.
\newblock \href {http://dx.doi.org/10.1088/0305-4470/38/13/001}
  {\path{doi:10.1088/0305-4470/38/13/001}}.

\bibitem{kampen68PhRv}
N.~G. {van Kampen}, {Relativistic Thermodynamics of Moving Systems}, Physical
  Review 173 (1968) 295--301.
\newblock \href {http://dx.doi.org/10.1103/PhysRev.173.295}
  {\path{doi:10.1103/PhysRev.173.295}}.

\bibitem{Israel76AnPhy}
W.~{Israel}, {Nonstationary irreversible thermodynamics: A causal relativistic
  theory}, Annals of Physics 100 (1976) 310--331.
\newblock \href {http://dx.doi.org/10.1016/0003-4916(76)90064-6}
  {\path{doi:10.1016/0003-4916(76)90064-6}}.

\bibitem{yuen70AmJPh}
C.~K. {Yuen}, {Lorentz Transformation of Thermodynamic Quantities}, American
  Journal of Physics 38 (1970) 246--252.
\newblock \href {http://dx.doi.org/10.1119/1.1976295}
  {\path{doi:10.1119/1.1976295}}.

\bibitem{tadas08arXiv}
T.~K. {Nakamura}, {Three Views of a Secret in Relativistic Thermodynamics}\href
  {http://arxiv.org/abs/0812.3725} {\path{arXiv:0812.3725}}.

\bibitem{jaynes83}
R.~D. Rozenkrantz (Ed.), Papers on probability statistics and statistical
  physics, Kluwer: Dordrecht, 1983.

\bibitem{dunkel07}
J.~{Dunkel}, P.~{Talkner}, P.~{H{\"a}nggi}, {Relative entropy, Haar measures
  and relativistic canonical velocity distributions}, New Journal of Physics 9
  (2007) 144--+.
\newblock \href {http://arxiv.org/abs/cond-mat/0610045}
  {\path{arXiv:cond-mat/0610045}}, \href
  {http://dx.doi.org/10.1088/1367-2630/9/5/144}
  {\path{doi:10.1088/1367-2630/9/5/144}}.

\bibitem{tadas09rel-eq}
T.~K. {Nakamura}, {Relativistic Equilibrium Distribution by Relative Entropy
  Maximization}\href {http://arxiv.org/abs/0909.2732} {\path{arXiv:0909.2732}}.

\end{thebibliography}

\end{document}